\documentclass[12pt,letterpaper]{article} 
\usepackage{amssymb}
\usepackage{amsmath}
\usepackage{amsfonts}
\usepackage{graphicx}
\usepackage{epstopdf}
\usepackage[]{hyperref}

\newenvironment{acknowledgments}{%
	\vskip 3.25ex 
	\addcontentsline{toc}{section}{Acknowledgments}
	\noindent {\bf Acknowledgments} 
}

\numberwithin{equation}{section}

\def\cf{{\it cf.}}

\newcommand{\beq}{\begin{equation}}
	\newcommand{\eeq}{\end{equation}}

\newcommand{\req}[1]{(\ref{#1})}

\begin{document}
	
	\begin{center}
	{\Large \bf On refined Chern-Simons / topological string duality  for classical gauge groups\\
		\vspace*{1 cm}
		
		{\large  M.Y. Avetisyan and R.L.Mkrtchyan
		}
		\vspace*{0.2 cm}
		
		{\small\it Yerevan Physics Institute, 2 Alikhanian Br. Str., 0036 Yerevan, Armenia}
		
	}
	
\end{center}\vspace{2cm}

{\small  {\bf Abstract.} 
	
We present the partition function of the refined Chern-Simons theory on $S^3$ with arbitrary A,B,C,D gauge algebra in terms of multiple sine functions. 
For B and C cases this representation is novel. It allows us to conjecture duality to some refined and orientifolded versions of the topological string on the resolved conifold, and carry out the detailed identification of different contributions.
The free energies for D and C algebras possess the usual halved contribution from the A theory, i.e. orientable surfaces, and contributions of non-orientable surfaces with one cross-cup, with opposite signs, similar as for the non-refined theories. However, in the refined case, both theories possess in addition a non-zero contribution of orientable surfaces with two cross-cups. In particular, we observe a trebling of the K\"ahler parameter, in the sense of a refinement and world-sheet (i.e. the number of cross-cups) dependent quantum shift. For B algebra the contribution of Klein bottles is zero, as is the case in the non-refined theory, and the one-cross-cup terms differ from the D and C cases. For the (refined) constant maps terms of these theories we suggest a modular-invariant representation, which leads to natural topological string interpretation.  We also calculate some non-perturbative corrections. 
	

\vskip 1em

\newpage 

\tableofcontents
	
\section{Introduction}

The concept of gauge/string duality has been at the forefront of developments in fundamental physics since its invention by 't Hooft \cite{tH}. One of the simplest cases where such a duality has been proven is the Chern-Simons/topological string duality. 

Witten's work showed that Chern-Simons theory is a completely solvable gauge theory with Wilson loops identifiable with knots \cite{W1}. The dual theory has been identified by Gopakumar and Vafa by showing that $SU(N)$ Chern-Simons theory on a 3-sphere can be interpreted at large N to be dual to the closed topological string on the resolved conifold \cite{GV1,GV2,GV3}. Not long after, it has been shown that for the other classical gauge groups ($SO,Sp$) the dual large N theory corresponds to an orientifold of the topological string on the resolved conifold \cite{SV}. In particular, these works showed that the perturbative partition functions of both theories match for a suitable identification of parameters.

Nekrasov's instanton solution of $N=2$ supersymmetric Yang-Mills theory, and its duality to M-theory, led to the development of a notion of a refined topological string \cite{N02,IKV07}. Here, refinement refers to the fact that the theory possesses besides the usual string coupling and K\"ahler parameters, a novel additional parameter. 

In their seminal work Aganagic and Shakirov were able to construct a novel refined Chern-Simons theory \cite{AS11,AS12a}, which connects to Macdonald's deformation of characters and related objects of simple Lie algebras \cite{Mac1,Mac2,Mac3}, as Macdonald polynomials play a major role in their refined theory. They also showed that for $SU(N)$ gauge group the refined Chern-Simons theory corresponds at large $N$ to the refined topological string on the resolved conifold. Subsequently, the case of $SO(2n)$ refined Chern-Simons was argued to be connected to the refined and orientifolded topological string on the resolved conifold \cite{AS12}. 

Somewhat orthogonal to these later developments, an universal expression for the Chern-Simons partition functions on $S^3$ has been proposed in \cite{MV,M13}, based on the notion of Universal Lie Algebra of Vogel \cite{V0,V}. This expression substitutes Witten's original representation of partition functions in terms of root system data by an integral representation, depending on the coupling constant and the homogeneous parameters of Vogel's projective plane.  

Surprisingly, it turned out that a slight generalization of the universal formulation also includes the refined Chern-Simons partition functions of Aganagic-Shakirov, as derived for $SU(N)$ and $SO(2n)$ groups in \cite{KS}, and later extended to the other groups in \cite{AM21}. One of the reasons why the universal formulation is of interest is because it allows to prove the gauge/string duality {\it exactly}, i.e., beyond perturbation theory. In particular, it has been shown in \cite{KM} that the duality for $SU(N)$ Chern-Simons on $S^3$ and the topological string on the resolved conifold holds non-perturbatively. Another outcome of the universal formulation is the suggestion that for exceptional gauge groups the corresponding Chern-Simons theories are actually dual to the refined closed topological string on some other backgrounds than the resolved conifold \cite{M21}.

The aim of the present paper is, first, to express the partition functions of B and C refined Chern-Simons theory on $S^3$, derived in \cite{AM21}, in terms of multiple sine functions. This requires a lengthy calculation, partially presented in Appendix \ref{secDerivBC}. This form is necessary to achieve our second aim, namely  to suggest the duality of refined Chern-Simons theory with refined topological strings for the remaining classical gauge groups $SO(2n+1), Sp(2n)$ (B, C type gauge algebras), as well as a deeper analysis of the duality for $SO(2n)$ (D type gauge algebra), already suggested first in \cite{AS12}. Another advantage of multiple sine form is that it contains non-perturbative corrections \cite{KM}, which we also derive below for D, C cases.

Our approach, as already mentioned, is based on a recently suggested compact form of the partition function of refined Chern-Simons theory on $S^3$ with an arbitrary gauge group \cite{AM21}, i.e., 

\begin{eqnarray}\label{refCS}
	Z_X(\kappa,y)= Vol(Q^{\vee})^{-1} \delta^{-\frac{r}{2}} \prod_{m=0}^{y-1} \prod_{\alpha_+} 2\sin \pi \frac{y(\alpha,\rho)-m \frac{(\alpha,\alpha)}{2}}{\delta}\,.
\end{eqnarray}

Here, $Vol(Q^{\vee})$ denotes the volume of the fundamental domain of the co-root lattice of the simple Lie algebra $X$, $r$ is the rank of the gauge group (simple Lie), $\alpha_+$ its positive roots, $y$ a refinement parameter assumed to be a positive integer ($y=1$ corresponds to the unrefined case), $\delta :=\kappa+y t$ with $\kappa$ the Chern-Simons coupling, and $t$ the group Coxeter number in an arbitrary normalization of the invariant scalar product in the algebra (becomes the true Coxeter number in the normalization with square of highest root equal to 2). 

Note that the partition function (\ref{refCS}) generalizes the partition functions of all known particular cases of Chern-Simons on $S^3$ \cite{AS11,AS12,AS12a,KS}. An important feature of (\ref{refCS}) is that
\begin{eqnarray}
	Z_X(0,y)=1\,.
\end{eqnarray}

The representation of $Z_X$ given in (\ref{refCS}) can be transformed into \cite{AM21,KS,KM}
\begin{equation}\label{Zky}
	Z_X(\kappa,y)= \left(\frac{ty}{\delta}\right)^{\frac{1}{2}dim(y)} Z_1 Z_2\,, 
\end{equation}
with
\begin{equation}
	\begin{split}
		\ln Z_1 &= -\int_{0}^{\infty} \frac{dx}{x(e^{x}-1)} \left( F_X\left( \frac{x}{\delta},y\right) - dim(y)\right)\,, \\
		\ln Z_2 &= \int_{0}^{\infty} \frac{dx}{x(e^{x}-1)} \left( F_X\left( \frac{x}{ty},y\right) - dim(y)\right)\,,
	\end{split}
\end{equation}
where $dim(y):=y(dim-r)+r$ and $dim$ being the dimension of the corresponding gauge algebra. $F_X(x,y)$ is the refined quantum dimension of the adjoint representation of the gauge algebra $X$, given by
\begin{eqnarray}\label{FX2}
	F_X(x,y)=    r+	\sum_{m=0}^{y-1}	\sum_{\alpha_{+}} \left( e^{x(y(\alpha,\rho) -m (\alpha,\alpha)/2)}+e^{-x(y(\alpha,\rho) -m (\alpha,\alpha)/2)}\right)\,.
\end{eqnarray}
In particular, $F_X(x,y)$ can be expressed as a ratio \cite{AM21}  
\begin{eqnarray} \label{FAB}
	F_X(x,y)= \frac{A_X}{B_X}\,,
\end{eqnarray}
with $A_X$ a polynomial of $q=\exp(x)$, and $B_X$ a product $B_X=(q^a-1)(q^b-1)\dots$ . The advantage of this representation is that it allows to express the partition function in terms of multiple-sine functions.

The general formula, which follows from (\ref{Zky})  and immediately leads to multiple-sine representation, is \cite{KM,AM21}
\begin{eqnarray}\label{FintRep}
	\ln Z=-\frac{1}{4}\int_{R_+} \frac{dx}{x} \frac{\sinh\left(x(ty-\delta)\right)}{\sinh\left( x ty \right)\sinh\left(x \delta\right)} F_X(2x,y)\,.
\end{eqnarray}
E.g. in the simplest case of unrefined $SU(N)$ Chern-Simons theory one obtains the partition function 

\begin{eqnarray} \label{SUN}
	Z_A=
	\frac{\sqrt{\delta}}{ \sqrt{N}}  \frac{S_3(1+N|1,1,\delta) }{S_3(1|1,1,\delta) }\,.
\end{eqnarray}
As has been shown earlier in \cite{SGV}, this is a very convenient representation to perform a further transformation into a Gopakumar-Vafa style BPS expansion. The BPS expansion can be used to establish a duality to the (refined) topological string. The duality has been established in a number of cases, but not for all.

The cases of the refined Chern-Simons theories with B, and C algebras, which are included in the classical gauge algebras, 
have been among the uncovered in this context.
Actually, as we shall see, for the known D case a more detailed duality
relation will be suggested, with an exact identification
(partially conjectured) of different parts of the partition function
through different kinds of string's surfaces.

The outline is as follows. In section \ref{secPartFunc} we will state the partition functions for refined Chern-Simons on $S^3$ with A,B,C,D type algebra in terms of multiple sine functions. As the derivation is rather lengthy and technical, we only state the final result and refer the reader to Appendices \ref{secMultSine} and \ref{secDerivBC} for further details about the calculations. In section \ref{CSTduality} we perform a Gopakumar-Vafa type expansion of the partition functions and discuss in detail matching of terms with  expected world-sheet contributions for different kinds of surfaces. Section \ref{secNPcorr} briefly discusses some non-perturbative aspects of unoriented contributions. We conclude in section \ref{secConclusion} with comparison with known duality of D case, and open problems.

\section{The partition functions of the refined Chern-Simons theory on $S^3$} \label{secPartFunc}

The partition functions of the refined Chern-Simons theory on $S^3$ for A,B,C and D gauge groups can be derived in terms of multiple sine functions using the expression (\ref{FintRep}) (cf. \cite{KM,AM21}). In detail, taking into account the expression (\ref{FAB}) for $F_X$, i.e. that its numerator can be represented as a sum of exponentials, we see that the  partition function (\ref{FintRep}) corresponds to a product of multiple sine functions $S_r$, defined by (see Appendix \ref{secMultSine} for more details):

\begin{equation} \label{multsine}
	\ln S_r(z|\underline{\omega})=
	(-1)^r \left( \frac{1}{2} \oint    \frac{dx}{x}\frac{e^{zx}}{\prod_{k=1}^r(e^{\omega_i x}-1)} +\int_{R_+}\frac{dx}{x}\frac{e^{zx}}{\prod_{k=1}^r(e^{\omega_i x}-1)}\right)\,.
\end{equation}
The partition function \req{FintRep} is a product of (powers of) different multiple sine functions up to fifth rank, as can be easily established from (\ref{FintRep}) and \req{multsine}. This expression can be simplified by making use of various identities of multiple sine functions, presented in Appendix \ref{secMultSine}. An outline of the derivation is given in Appendix \ref{secDerivBC}. We present below the final results, which we in addition verified by numerical checks (making use of $Mathematica^{\text{\textcopyright}}$). 

\paragraph{$A_{N-1}$}
The partition function of refined Chern-Simons on $S^3$ with $SU(N)$ gauge group (i.e., $A_{N-1}$ algebra) reads
\begin{equation} 
	\begin{split}
		Z_A(a,y,\delta) &= \left(\frac{ty}{\delta}\right)^{\frac{1}{2}dim(y)}
		\exp{\int_{0}^{\infty} \frac{dx}{x(e^{x}-1)} \left( F_A\left( \frac{x}{ty},y\right) - F_A\left( \frac{x}{\delta},y\right)\right) }\\ &=  \label{ZAr1}
		\sqrt{\frac{\delta}{ya}} \frac{S_3(1+ya|1,y,\delta)}{S_3(y|1,y,\delta)}\,,
	\end{split}
\end{equation}
where $a=t=N$ is (the inverse) parameter of the $1/N$ expansion. This expression has been already derived in \cite{KM}. We further verified this expression numerically for a random sample of points in the  $\delta, N, y$ parameter space. 

\paragraph{$D_{N/2}$}
The expression for $SO(N)$ algebras of type D (i.e., even $N$) has been originally derived in \cite{KM}, and can be written in terms of multiple sine functions as 
\begin{eqnarray} \label{ZDr1}
	Z_{D}= 
	\frac{1}{\sqrt{2}}    \frac{S_3(1+ay|1,2y,\delta)}{S_3(y|1,2y,\delta)\, S_2(2-y+ ya |2,2\delta)}\,,
\end{eqnarray}
with $a=N-1$. We numerically verified this expression as well.

\paragraph{$C_{N/2}$}

The partition function for $Sp(N)$ algebras (type C,  $N$ even) can be calculated using the expression for $F_C$ derived in \cite{AM21},  and reads
\begin{eqnarray} \label{ZCr1}
	Z_C =	\frac{S_3(1+ya|1,2y,2\delta)}{S_3(y|1,2y,2\delta)} \frac{S_2(2-y+ ya|2,4\delta)}{S_2(2-y+ ya|2,2\delta)}\,,
\end{eqnarray}
with $a=N+1$. The analytic derivation of the above $Z_C$ is far more involved than the previous cases, and is outlined in Appendix  \ref{secDerivBC}. Again, we numerically confirm our analytic derivation.

\paragraph{$B_{(N-1)/2}$}

Finally, for the remaining case of $SO(N)$ group with odd $N$, i.e., the B series, we derived 
\begin{equation}\label{ZBr1}
	\begin{split}
		Z_{B} =&\, 
		\frac{1}{2} \frac{S_3(1+ay|1,2y,\delta)}{S_3(1|1,2y,\delta)} \frac{1}{S_2(y(a+1)/2|y,\delta)} \\ &\times\frac{S_3(y+ay|1,2y,2\delta)}{S_3(1+ay|1,2y,2\delta)} 
		\frac{S_3(1|1,2y,2\delta)}{S_3(1+y|1,2y,2\delta)}\,,
	\end{split}
\end{equation}	
with $a=N-1$. As for the previous cases, we  numerically verified this result.

\section{Chern-Simons / topological string duality}
\label{CSTduality}

The Chern-Simons / topological string duality states that the partition function of Chern-Simons theory on a $3$-sphere is equivalent to the partition function of the topological string on a specific Calabi-Yau manifold, namely a resolved conifold, as originally put forward in \cite{GV1,GV2,GV3,OV02}. The duality can be extended to the refined case \cite{IKV07,N02}, as discussed more recently in \cite{AS11,AS12,AS12a,KS} for $SU(N)$ and $SO(N)$. In particular, it has been shown in \cite{KM} that for non-refined $SU(N)$ the duality holds non-perturbatively, and is therefore exact.   

In this section we will derive from the partition functions of B, C, D types, given in the previous section, the  Gopakumar-Vafa type expansions. This form of the partition functions is  most suitable for a (refined) topological string interpretation. In particular, we propose that the well-known quantum shift of the K\"ahler parameter \cite{KW10a,KW10b} now depends on world-sheet surface types, i.e., is different for surfaces with zero, one, or two cross-cups. We will refer to this effect as a {\it trebling} of the K\"ahler parameter. 

\subsection{Partition functions reformulation}

The expressions for D and C partition functions most suitable for a topological string interpretation can be obtained from the initial expressions given in equations \req{ZDr1} and \req{ZCr1} by multiple sines manipulations, and read (again, we verified  results numerically): 
\begin{eqnarray} \label{ZDr}
	Z_D= \frac{1}{\sqrt{2}}	
	\frac{\sqrt{S_3(1+ya|1,y,\delta)}}{\sqrt{S_3(y|1,y,\delta)}}\times \\  \label{D-KB}
	\frac{\sqrt{S_3(y|1,y,\delta)}}{S_3(y|1,2y,\delta)}
	\frac{S_3(2+ya|1,2y,\delta)S_2(1+ya|2y,\delta)}{\sqrt{S_3(1+ya|1,y,\delta)}\sqrt{S_2(2-y+ya|2,\delta)}} \times \\  \label{D-unor}
	\frac{\sqrt{S_2(2-y+ya|2,\delta)}}{S_2(2-y+ya|2,2\delta)}\,,
\end{eqnarray}

\begin{eqnarray}\label{ZCr}
	Z_C= 	
	\frac{\sqrt{S_3(1+ya|1,y,\bar{\delta})}}{\sqrt{S_3(y|1,y,\bar{\delta})}}\times \\ \label{C-KB}
	\frac{\sqrt{S_3(y|1,y,\bar{\delta})}}{S_3(y|1,2y,\bar{\delta})}
	\frac{S_3(2+ya|1,2y,\bar{\delta})S_2(1+ya|2y,\bar{\delta})}{\sqrt{S_3(1+ya|1,y,\bar{\delta})}\sqrt{S_2(2-y+ya|2,\bar{\delta})}} \times \\ \label{C-unor}
	\frac{S_2(2-y+ya|2,2\bar{\delta})}{\sqrt{S_2(2-y+ya|2,\bar{\delta})}}\,,
\end{eqnarray}
where in the C case we introduced $\bar{\delta}:=2\delta$.  

These formulae are quite remarkable. We see that the answers for D and C almost coincide. Only the last fractions 
(\ref{D-unor}) and (\ref{C-unor}) are different, more precisely, they are inverse to each other. The same feature is observed in the non-refined limit, i.e., at $y=1$. In this limit, the above expressions reduce to (see  \cite{SGV,LV12}): 
\begin{eqnarray} \label{SON}
	Z_{D}(y=1)=2^{-\frac{3}{4}} \sqrt{\frac{S_3(1+a|1,1,\delta)}{S_3(1|1,1,\delta)}} \frac{\sqrt{S_2(1+a|2,\delta)}}{S_2(1+a|2,2\delta)}\,, 
\end{eqnarray}
and
\begin{eqnarray} \label{SpN}
	Z_C(y=1)=
	2^{-\frac{1}{4}}\sqrt{\frac{S_3(1+a|1,1,\bar{\delta})}{S_3(1|1,1,\bar{\delta})}} \frac{S_2(1+a|2,2\bar{\delta})}{\sqrt{S_2(1+a|2,\bar{\delta})}}\,.
\end{eqnarray}
For simplicity, from now on we will omit in formulae for C the bar over $\delta$. Note that in the non-refined limit the last fraction can be interpreted  as a contribution of one-cross-cup non-orientable surfaces to the dual topological string partition function \cite{SV}. Hence,  their contributions to the free energy in the D and C cases are the same, up to a sign.
As we shall see soon, the same interpretation appears to be suggestive for the refined case, with again D and C theories to differ simply by the sign of the contribution of the one-cross-cup non-orientable surfaces. It is remarkable that despite the complicated derivation of the refined formulae, such a simple relation between the D and C theories persists. 

Let us also rewrite the B theory in a form which will be advantageous for later derivations:
\begin{eqnarray} \label{ZBr}
	Z_{B}=	 
	\frac{1}{2}
	\frac{\sqrt{S_3(1+ya|1,y,\delta)}}{\sqrt{S_3(y|1,y,\delta)}}\times \\ \label{ZBconst} 
	\frac{\sqrt{S_3(y|1,y,\delta)}}{S_3(1|1,2y,\delta)} \frac{S_3(1|1,2y,2\delta)}{S_3(1+y|1,2y,2\delta)}\times \\  \label{B-unor}
	\frac{\sqrt{S_3(1+ya+\delta|1,2y,2\delta)}}{\sqrt{S_3(1+ya|1,2y,2\delta)}} \frac{\sqrt{S_3(y+ay+1|1,2y,2\delta)}}{\sqrt{S_3(1+y+ay+\delta|1,2y,2\delta)}} \,.
\end{eqnarray}	
Note that while in the non-refined
limit ($y=1$) the B and D theories coincide, this does not hold anymore for the refined theories. 

\subsection{Gopakumar-Vafa expansions}

In order to faciliate an interpretation of the partition functions (\ref{ZDr}) and (\ref{ZCr}) at $y\neq 1$ in terms of the refined topological string theory, we transform the {\it perturbative} part of the sines into the Gopakumar-Vafa type expansions of the topological string theory partition functions, following the recipe given in \cite{SGV,KM}. In detail, by closing the integration contour in the upper semiplane in eq. (\ref{multsine}) we can express the integral as a sum of residues of the poles in the upper semiplane. The distinguished poles are those coming from the term $(e^x-1)$ in the denominator. Introducing the string coupling constant $g_s=2\pi i/\delta$ we can see that these poles yield the perturbative contributions over $g_s$, while the remaining poles yield non-perturbative terms $\sim \exp{(-1/g_s)}$, \cf, \cite{SGV,KM}. Note that in addition to Gopakumar-Vafa contributions topological strings have also contributions from constant maps. Such contributions come from the multiple sines in partition functions not dependent on parameters $a$, i.e.,   on K\"ahler parameters, see below. 
However, the similar approach, separating the "perturbative" parts of constant maps, leads to the incomplete and/or incorrect results. We propose below a correct method for handling constant maps. 

{\bf D and C theories}

The first fraction in the D (\ref{ZDr}) and C (\ref{ZCr}) partition functions corresponds to the square root of the A type refined partition 
function (\ref{ZAr1}). This is in agreement with an orientifold interpretation. The corresponding perturbative part of the free energy reads 
\begin{eqnarray} \label{GVSU}
	\mathcal F_A = \frac{1}{2}  \sum_{n=1}^{\infty}\frac{e^{\frac{g_s}{2}n(1+2a y-y)}}{n(e^{\frac{g_s}{2} n}-e^{-\frac{g_s}{2} n})(e^{\frac{g_s}{2} n y}-e^{-\frac{g_s}{2} n y})} =
	\frac{1}{2}  \sum_{n=1}^{\infty}\frac{e^{\tau_0 n}}{n(q^{ n}-q^{- n})(t^{ n}-t^{-n})}\,,
\end{eqnarray}
with
\begin{equation}\label{qshift}
	q=e^{\frac{g_s}{2}}\,,\,\,\,\,\,t=e^{\frac{g_s}{2} y} \,,\,\,\,\,\, \tau_0 =\frac{g_s}{2} (1+2a y-y)=g_s a y+\frac{g_s}{2}(1-y)\,,
\end{equation}
which is the well-known refined (BPS) version of the Gopakumar-Vafa free energy of the refined topological string on the resolved conifold, \cf, \cite{IKV07}. This is dual to the A type refined Chern-Simons theory. 

An important feature of (\ref{GVSU}) is the quantum shift of the K\"ahler parameter $\tau_0$ given in (\ref{qshift}). At $y=1$ $\tau_0$ turns into the classical K\"ahler parameter $\tau=a g_s$. We shall see that such quantum shifts take also place for the other parts of the partition functions of the B, C, D cases, with a novel and interesting detail – the quantum shifts differ for different types of world-sheet surfaces. 

Next, consider the last fractions (\ref{D-unor}) and (\ref{C-unor}. They coincide with those in the non-refined case, 
and can be interpreted as the contribution of non-orientable one-cross-cup surfaces \cite{SV}. However, the corresponding K\"ahler parameter is now shifted, with a different shift than for the oriented surfaces. In detail,
the perturbative contribution for D type reads
\begin{eqnarray}\label{Dc=1}
	\sum_{n=1,3,...} \frac{1}{n} \frac{e^{\frac{1}{2} \tau_1 n}}{(q^n-q^{-n})}\,, \\
	\tau_1=  g_s(1+ay-y)\,,
\end{eqnarray}
and identically with opposite overall sign for the C case. This expression is odd over the string coupling, provided $\tau_1$ is invariant. This agrees with an interpretation as a contribution of non-orientable one cross-cup surfaces, since the power of coupling constant is the Euler characteristic, and the Euler characteristic of such surfaces is odd. 

The remaining, coinciding, terms in the D and C partition functions (\ref{D-KB}), (\ref{C-KB}) can be rewritten as
\begin{eqnarray}\label{c=2}
	\sum \frac{1}{2n} \frac{\frac{q^n}{t^n}-\frac{t^n}{q^n}}{(q^{2n}-q^{-2n})(t^{n}+t^{-n})} e^{\tau_2 n} \\
	\tau_2= (1+2ay-2y) \frac{g_s}{2}
\end{eqnarray}
This expression is even over $g_s$, provided $\tau_2$ is invariant, and can therefore be interpreted as a contribution of a Klein bottle with handles,
since the Euler characteristics of such surfaces is even. Note that in the non-refined case there is no such contribution \cite{SV}, and in accordance this expression becomes zero at $y=1$. 

In summary, we observe that the single K\"ahler parameter $\tau= a g_s$ of the non-refined case appears to treble into three parameters after refinement, i.e.,
\begin{eqnarray} \label{tau0}
	\tau_0 =  (1+2ay-y) \frac{g_s}{2} =y \tau +\frac{1}{2}(1-y) g_s  \\ \label{tau1}
	\tau_1=  (1+ay-y)g_s  = y \tau +(1-y) g_s\\ \label{tau2}
	\tau_2= (1+2ay-2y) \frac{g_s}{2} = y \tau +\frac{1}{2}(1-2y) g_s \\
	\tau = a g_s
\end{eqnarray}

Note that $\tau_0, \tau_1$ become $\tau$ at unrefined limit $y=1$, but in the same limit  $\tau_2=\tau-\frac{1}{2}g_s$, i.e. doesn't coincide with initial K\"ahler parameter $\tau$. However, the coefficient in front of $exp(n\tau_2)$ in contribution (\ref{c=2}) of two cross-cups surfaces tends to zero, so altogether (\ref{c=2}) tends to zero, as it should, since there is no such contribution in the unrefined limit. 

{\bf B theory}

Finally, we shall carry out a similar analysis for the B type theory. We have the same refined A type terms, which correspond to the contributions of Riemannian surfaces, i.e., orientable surfaces with handles. These have even Euler characteristics. As before, we ignore the constant terms (\ref{ZBconst}) and consider the remaining contribution (\ref{B-unor}), whose GV type expansion can be calculated to be given by
\begin{eqnarray}\label{Bc=1}
	\sum_{n=1,3,...} \frac{1}{n} \frac{e^{\frac{1}{2} \tau_1^B n}}{(q^{n/2}-q^{-n/2})(t^{n/2}+t^{-n/2})}\,, \\
	\tau_1^B = g_s (y a- y \frac{1}{2}+\frac{1}{2})= y\tau +\frac{1}{2}(1-y) g_s\,.
\end{eqnarray}

This is the deformed version of the one-cross-cup term of the non-refined $SO$ theory. It is odd w.r.t. the string coupling $g_s$, provided $\tau_1^B$ is unchanged. 
Note that the deformation $\tau_1^B$ is different from $\tau_1$ in the D (and C) cases. We conclude that in the B case there is no Klein bottle contribution, as in the non-refined case.
The initial K\"ahler parameter $\tau$ is shifted for Riemannian surfaces and for unorientable one cross-cup surfaces as follows:

\begin{eqnarray} \label{tauB}
	\tau_0^B =  (1+2ay-y) \frac{g_s}{2} =y \tau +\frac{1}{2}(1-y) g_s\,,  \\ 
	\tau_1^B = g_s (y a- y \frac{1}{2}+\frac{1}{2})= y\tau +\frac{1}{2}(1-y) g_s\,, \\
	\tau = a g_s\,.
\end{eqnarray}
We conclude that for the B case the shifts are identical, i.e., $\tau_0^B= \tau_1^B$. Taking into account the absence of Klein bottles contribution, we see that refined B case is very close to the unrefined D case.  Main difference is in the one cross-cup contribution (\ref{Bc=1}), in the presence of  $(t^{n/2}+t^{-n/2})$ contribution in the denominator. 

\subsection{Refined constant maps}

The constant maps contributions in the above partition functions
correspond to the multiple sine multipliers, which do not
depend on the parameter $a$, i.e. those independent on the 
K\"ahler parameter $\tau=a g_s/2$.
In case of the D theory there are constant maps, coinciding with those
in the A theory part (\ref{ZDr}) and for the 
(as we shall see) Klein bottle contribution (\ref{D-KB}):

\begin{eqnarray} \label{DConst}
	\frac{\sqrt{S_3(y|1,y,\delta)}}{S_3(y|1,2y,\delta)}
\end{eqnarray}

The perturbative poles contribution in this
expression is as follows: 

\begin{eqnarray} \label{s212r}
	\frac{1}{2} \sum_{n=1}^{\infty} \frac{(\frac{t}{q})^n}{n(q^n-q^{-n})(t^n+t^{-n})}
\end{eqnarray}

This term is neither even, nor odd over the 
coupling constant $g_s$ ($g_s \rightarrow {-g_s}$ is equivalent 
to $q \rightarrow {1/q}, t \rightarrow {1/t}$).
Another drawback of this expression reveals itself at $y=1$ ($q=t$).
It remains non-zero in this limit: 

\begin{eqnarray} \label{s212}
	\frac{1}{2} \sum_{n=1}^{\infty} \frac{1}{n(q^{2n}-q^{-2n})}
\end{eqnarray}

However, the initial expression is 

\begin{eqnarray}
	\frac{\sqrt{S_3(y|1,y,\delta)}}{S_3(y|1,2y,\delta)} = \frac{1}{\sqrt{S_2(1|2,\delta)}}= 2^{-\frac{1}{4}}
\end{eqnarray}
so its expansion should be zero. 
Resolution of this discrepancy is the following:
(\ref{s212}) is indeed the perturbative poles expansion of
$S_2(1|2,\delta)$, but one has to take into account the remaining
non-perturbative poles, coming from the second parameter, $2$. These two contributions cancel 
due to their modular covariance properties (i.e. properties under $g_s \rightarrow 1/g_s$), providing final result  according to $S_2(1|2,\delta)=\sqrt{2}$.

To handle this ambiguity, we suggest rewriting all terms corresponding
to the constant maps in an identical "modularity-neutral" form, i.e.
as follows:

\begin{equation}
	S_r(z|w)=\sqrt{S_r^2}=\sqrt{S_r(z|w)S_r(|w|-z|w)^{(-1)^{r+1}}}
\end{equation}

In this case, $S_2(1|2,\delta)$ will not contribute in the
GV expression. Indeed, the contributions of perturbative poles of the
two $S_2$ terms coming from $S_2(1|2,\delta)$ are:

\begin{eqnarray}
	\ln S_2(1|2,\delta) \sim \sum_{n=1}^\infty \frac{1}{n} \frac{e^{g_s n}}{(e^ {2g_s n}-1) }\\
	-\ln S_2(1+\delta|2,\delta) \sim -\sum_{n=1}^\infty \frac{1}{n} \frac{e^{g_s n}}{(e^ {2g_s n}-1) }
\end{eqnarray}
which cancel each other out.

The refined constant maps term in (\ref{SUN}), (\ref{SON}),
corresponding to the
Riemann surfaces (i.e. orientable worldsheets),
in GV form will contribute by the following expression:

\begin{eqnarray} \label{Aconst}
	\ln	S_3(y|1,y,\delta)= \ln \sqrt{S_3(y|1,y,\delta)S_3(1+\delta|1,y,\delta)} \sim \\ 
	-\frac{1}{2} \sum_{n=1}^\infty \frac{1}{n} \frac{e^{g_s y n}}{(e^ {g_s n}-1)(e^ {g_s y n}-1) } 
	-\frac{1}{2} \sum_{n=1}^\infty \frac{1}{n} \frac{e^{g_s n}}{(e^ {g_s n}-1)(e^ {g_s y n}-1) }= \\
	-\frac{1}{2}	\sum \frac{1}{n} \frac{\frac{q^n}{t^n}+\frac{t^n}{q^n}}{(q^n-q^{-n})(t^n-t^{-n})} \\
	q=e^{\frac{g_s}{2}},\,\,\,t=e^{\frac{g_s y}{2}}
\end{eqnarray}

This expression has the same limit in unrefined case at $y=1$, however,
it is explicitly even w.r.t. the string coupling $g_s$. 

Without our recipe, the initial constant maps term  $S_3(y|1,y,\delta)$  of refined A theory leads to the expansion \cite{KM}

\begin{eqnarray} \label{FAPconifold}
	\frac{1}{2}\sum_{n=1}^\infty\frac{ e^{- n g_s (1-y)/2}}{n \sinh\left(\frac{n g_s}{2}\right)\sinh\left(\frac{n y g_s}{2}\right)}\,,
\end{eqnarray}

To be even over $g_s$ it requires simultaneous change of $y$, hence uses an overall invariance w.r.t. the transformation $y \rightarrow 1/y, g_s\rightarrow y g_s $, which is the feature of the $A$ theory, only. 

One can present the product form of the \req{Aconst}, 
to compare it with MacMahon (or refined MacMahon \cite{IKV07})
function:

\begin{eqnarray}
	\exp \left(	\sum \frac{1}{n} \frac{\frac{q^n}{t^n}+\frac{t^n}{q^n}}{(q^n-q^{-n})(t^n-t^{-n})} \right) = \\
	\prod_{i,j=1}^{\infty} \frac{1}{(1-q^{i+1}t^{j-1})(1-q^{i-1}t^{j+1})}
\end{eqnarray}
which appear to be a $q  \leftrightarrow t$ symmetrized version of that in \cite{IKV07}. 

Next, consider the new form of the constant maps term for the
refined D theory (\ref{DConst}):

\begin{eqnarray}
	\ln	\frac{\sqrt{S_3(y|1,y,\delta)}}{S_3(y|1,2y,\delta)} \sim \\
	\ln 	\frac{\sqrt{\sqrt{S_3(y|1,y,\delta)S_3(1+\delta|1,y,\delta)}}}{\sqrt{ S_3(y|1,2y,\delta) S_3(1+y+\delta|1,2y,\delta)}}   \sim  \\
	- \frac{1}{4}	\sum \frac{1}{n} \frac{\frac{q^n}{t^n}-\frac{t^n}{q^n}}{(q^n-q^{-n})(t^{n}+t^{-n})}
\end{eqnarray}

Remarkably, this expression is even in $g_s$, which allows
us to assume that it corresponds to the constant maps term of a
two cross-cup non-orientable worldsheet contribution. 

Consider remaining constant maps terms, namely those in B case, two fractions in \req{ZBconst}. They are 

\begin{eqnarray}	
	\ln	\frac{\sqrt{S_3(y|1,y,\delta)}}{S_3(1|1,2y,\delta)} \sim \sum_{n=1} \frac{\frac{t^n}{q^n}-\frac{q^n}{t^n}}{4 n \left(q^n-q^{-n}\right) \left(t^{-n}+t^n\right)} \\
	\ln	\frac{S_3(1|1,2y,2\delta)}{S_3(1+y|1,2y,2\delta)}  \sim - \sum_{n=1} \frac{\frac{t^{n/2}}{q^{n/2}}-\frac{q^{n/2}}{t^{n/2}}}{2 n \left(q^{n/2}-q^{-\frac{n}{2}}\right) \left(t^{-\frac{n}{2}}+t^{\frac{n}{2}}\right)}
\end{eqnarray}	
so the sum of these two terms is the sum over odd $n$ :

\begin{eqnarray}	
	\ln \frac{\sqrt{S_3(y|1,y,\delta)}}{S_3(1|1,2y,\delta)}	\frac{S_3(1|1,2y,2\delta)}{S_3(1+y|1,2y,2\delta)}  \sim \\
	- \sum_{n=1,3,} \frac{\frac{t^{n/2}}{q^{n/2}}-\frac{q^{n/2}}{t^{n/2}}}{2 n \left(q^{n/2}-q^{-\frac{n}{2}}\right) \left(t^{\frac{n}{2}}+t^{-\frac{n}{2}}\right)}
\end{eqnarray}

We see, that answer is even w.r.t. the string coupling, so we assume that it again is coming from Klein bottles, representing their refined constant maps contribution on orientifolded resolved conifold. At $y=1$ this expression is zero.

\section{Non-perturbative corrections}
\label{secNPcorr}
Non-perturbative corrections to the D theory were derived in \cite{KM}. In principle, non-perturbative corrections are not obliged to be connected with the different types of world-sheet surfaces, which contribute to the perturbative (Gopakumar-Vafa) expansion of the free energy. However, in our case they are. The reason being that non-perturbative corrections are coming from poles other than $1/(e^x-1)$ of each sine function. But sine functions in the partition functions are distributed, according to above analysis, between different types of surfaces, and so naturally the non-perturbative parts can be associated with the type of surface.  This, in turn,  means that in each of these non-perturbative contributions to the refined partition function we should use the corresponding shifted K\"ahler parameter. This implies a corresponding rewriting of the formulae of \cite{KM}. The same expression holds also for the C case with an appropriate sign change in the front of the one-cross-cup contribution. For the B case one has to calculate one additional term corresponding to the new one-cup non-orientable surfaces. 

The explicit calculation of non-perturbative corrections for different types of surfaces is straightforward. The resulting expression are however lengthy, and at this moment there is no independent calculations for surfaces with one or two cross-cups. Below we present explicit expressions for the Klein-bottle surfaces for the D (and C) cases.

The constribution of non-perturbative poles for the D and C cases, i.e., obtained from expression (\ref{D-KB}) without the first fraction, which corresponds to constant maps terms, is given by the following expression
\begin{eqnarray}
	\sum_{n=1}^\infty \frac{1}{2n} \frac{K_1}{K_2}\,,
\end{eqnarray}
where
\begin{eqnarray}
	K_2=\left(-1+e^{2 \pi  i \delta  n}\right) \left(1+e^{2 \pi  i n y}\right) \left(-1+e^{\frac{2 \pi  i n}{y}}\right) \left(-1+e^{\frac{2 \pi  i \delta  n}{y}}\right)\,,
\end{eqnarray}
and
\begin{eqnarray}
	\begin{split}
		K_1 =& -e^{i (a-1) n \pi  y}+e^{2 i a n \pi  y}-e^{i (a+1) n \pi  y}+2 e^{\frac{i n \pi  (a y+1)}{y}}-e^{\frac{2 i n \pi  (a y+1)}{y}}\\&+2 e^{\frac{i n \pi  (a y+2)}{y}}-e^{\frac{2 i n \pi  (y^2+a y+1)}{y}}+2 e^{\frac{i n \pi  (2 y^2+a y+1)}{y}}+2 e^{\frac{i n \pi  (2 y^2+a y+2)}{y}}-e^{\frac{2 i n \pi  (a y^2+1)}{y}}\\&+e^{\frac{i n \pi (a y^2-y^2+2)}{y}}+e^{\frac{i n \pi  (a y^2+y^2+2)}{y}}+2 e^{\frac{i n \pi  (a y+\delta +1)}{y}}+2 e^{\frac{i n \pi  (a y+\delta +2)}{y}}-e^{i n \pi  (a y-y+\delta )}\\&-e^{i n \pi  (a y+y+\delta )}+2 e^{\frac{i n \pi (2 y^2+a y+\delta +1)}{y}}+2 e^{\frac{i n \pi  (2 y^2+a y+\delta +2)}{y}}-e^{\frac{2 i n \pi (a y^2+\delta )}{y}}+e^{\frac{2 i n \pi (a y^2+\delta +1)}{y}}\\&+e^{\frac{i n \pi (a y^2-y^2+2 \delta )}{y}}-e^{\frac{i n \pi (a y^2-y^2+2 \delta +2 )}{y}}+e^{\frac{i n \pi (a y^2+y^2+2 \delta)}{y}}-e^{\frac{i n \pi (a y^2+y^2+2 \delta +2)}{y}}\\&+e^{\frac{2 i n \pi  (a y+\delta  y+1)}{y}}+e^{\frac{2 i n \pi  (y^2+a y+\delta  y+1)}{y}}+e^{\frac{i n \pi  (a y^2-y^2+\delta  y+2)}{y}}+e^{\frac{i n \pi  (a y^2+y^2+\delta  y+2)}{y}}\\&+e^{\frac{i n \pi  (a y^2-y^2+\delta  y+2 \delta )}{y}}-e^{\frac{i n \pi (a y^2-y^2+\delta  y+2 \delta +2)}{y}}+e^{\frac{i n \pi  (a y^2+y^2+\delta  y+2 \delta )}{y}}-e^{\frac{i n \pi (a y^2+y^2+\delta  y+2 \delta +2)}{y}}\\&-2 e^{\frac{i n \pi  (a y+2 \delta  y+1)}{y}}-2 e^{\frac{i n \pi  (a y+2 \delta  y+2)}{y}}-2 e^{\frac{i n \pi  (2 y^2+a y+2 \delta  y+1)}{y}}-2 e^{\frac{i n \pi  (2 y^2+a y+2 \delta  y+2)}{y}}\\&-2 e^{\frac{i n \pi  (a y+2 \delta  y+\delta +1)}{y}}-2 e^{\frac{i n \pi  (a y+2 \delta  y+\delta +2)}{y}}-2 e^{\frac{i n \pi (2 y^2+a y+2 \delta  y+\delta +1)}{y}}-2 e^{\frac{i n \pi (2 y^2+a y+2 \delta  y+\delta +2)}{y}}\,.
	\end{split}
\end{eqnarray}

Expressing $\delta$ in terms of the string coupling $\delta = 2\pi i /g_s$, we evidently get non-perturbative contributions of type $exp(-1/g_s)$. (We did not explicitly substitute in the above in order to avoid further complication of the lengthy expression.) Similarly, one can introduce a K\"ahler parameter $\tau_2$ by expressing $a$ in terms of $\tau_2$, i.e., $a=(2\tau_2/g_s+2y-1)/(2y)$. 

Note that similar calculations of non-perturbative corrections, carried out in \cite{KM,SGV}, coincide with purely string calculations \cite{HMMO13,H15}. For the Klein bottle contributions derived here, there is presently no corresponding calculation from topological string theory available to compare with.

\section{Conclusion}
\label{secConclusion}

In this work we derived exact expressions for the partitions functions of the refined Chern-Simons theories of B, C types (A, D were known earlier, \cf, \cite{AS11,AS12,AS12a,KS,KM}) on $S^3$ in terms of multiple sine functions. Such expressions are convenient for investigation of their dualities with topological strings, and for derivation of non-perturbative corrections. Although the derivation is lengthy (partially presented in Appendix \ref{secDerivBC}), the answers are simple and very close to each other.

Then, following the now standard procedure, we derived a Gopakumar-Vafa type representation of the free energies (i.e. separate the perturbative, over string coupling constant, part), and suggest an identification of the contributions with different types of world-sheet surfaces.  We observed the trebling phenomenon. It means that in the case of the refined theory the shifts of the initial K\"ahler parameter are different for three types of surfaces, contributing into partition function. 
Two of these parameters can be described as deformations of the initial one, so that in the non-refined limit, they tend to the initial K\"ahler parameter. The third one, namely that for the Klein bottles, is not equal to the initial one at $y=1$. However, having that the coefficients corresponding to this surface becomes zero in that limit, this third parameter  does not contribute into the non-refined limit. 

This interpretation should be stressed to be a hypothesis – a natural one, though still to be confirmed by some independent string-theory calculations of the refined amplitudes, connected with different orientifolded surfaces. Such calculations are yet to be carried out.   
However, this interpretation in terms of different types of surfaces itself supports the idea of an existence of some microscopic theory of the refined topological strings, which is still unknown (although consider \cite{AAFJ}). The expansion over the string coupling $g_s$ still satisfies the rules of the non-refined case – the powers $g^\chi$ are given by the Euler characteristics $\chi$ of the corresponding surfaces. However, there is no path integral representation for the coefficients of this expansion, which now depend on the additional refinement parameter $y$.  

The results for D case can be compared with those in \cite{AS12}, where the GV type partition function for D case was calculated for the first time, and a duality with the refined topological strings was stated. 
The GV type part of our free energy for D case (i.e., the sum of (\ref{GVSU}), (\ref{Dc=1}), and (\ref{c=2})) coincides with their result ((7,8) or (7.9) in \cite{AS12}), after a redefinition of variables. 
The difference is that here we suggested a more detailed correspondence of terms of GV type expansion with different types of surfaces, similar to that suggested in \cite{SV} 
in the non-refined limit $y=1$. Of course, our answers in the non-refined limit agree with those in \cite{SV}.

This study can be continued in several directions. In  \cite{AM22} (in preparation) we hope to address the limit of the Chern-Simons theories near the conifold singularity.
In this limit the corresponding partition functions appear to be coinciding with the refined version of the universal volume of the groups. This slightly differs from the previously studied refined matrix model \cite{ABCD-K}, it is interesting to compare conclusions of these two models. 

Another possible direction of research is the extension of the approach presented on the exceptional gauge groups. In case of these groups the non-refined Chern-Simons theory is suggested \cite{M21} to be dual to a refined topological string. However, the problem of refined version of the Chern-Simons theories for the exceptional groups remains open.

\acknowledgments{
	
	We are indebted to D.Krefl for discussions of the present work. 	
	The work of MA was fulfilled within the Regional Doctoral Program on Theoretical and Experimental Particle Physics  sponsored by VolkswagenStiftung. MA and RM are partially supported by the Science Committee of the Ministry of Science and Education of the Republic of Armenia under contracts 20AA-1C008, 21AG-1C060. The work of MA is partially supported by ANSEF grant PS-mathph-2697. 
}

\appendix
\section{Multiple sine functions}
\label{secMultSine}
For the readers convenience, we will recall in this appendix some basic properties of the multiple sine functions $S_r(z|\underline{\omega})$ of importance for the derivations in this work, see \cite{NARU03,KK03}.

The integral representation of multiple sine functions reads
\begin{equation}\label{Sint}
	\begin{split}
		\ln S_r(z|\underline{\omega})&=
		(-1)^r \frac{1}{2} \oint    \frac{dx}{x}	I(x,z|\underline{\omega}) +(-1)^r \int_{R_+}\frac{dx}{x}	I(x,z|\underline{\omega})\\ &=
		(-1)^{r-1} \frac{1}{2} 	I(x,z|\underline{\omega})+(-1)^r \int_{R_-}\frac{dx}{x}	I(x,z|\underline{\omega})\,,
	\end{split}
\end{equation}
with
$$
I(x,z|\underline{\omega})=\frac{e^{zx}}{\prod_{k=1}^r(e^{\omega_i x}-1)}\,,
$$
and where the integration over the small circle at the origin proceeds counterclockwise. 

Some basic properties of $S_r(z|\underline{\omega})$ useful for analytic manipulations:
\begin{eqnarray}
	S_r(z|a_1,...,a_r)= S_r(\underline{a}-z|a)^{(-1)^{r+1}}\,, \\
	\underline{a}=a_1+...+a_r \,,\\
	S_r(cz|ca_1,...,ca_r)= S_r(z|a_1,...,a_r)\,, \\
	S_r(z+a_i|a_1,...,a_r)= 	\frac{S_r(z+a_i|a_1,...,a_r)}{S_r(z+a_i|a_1,...,a_{i-1},a_{i+1},...,a_r} \,,\\ \label{fusion}
	S_r(z+a|2a,a_2...,a_r) S_r(z|2a,a_2...,a_r)=S_r(z|a,a_2...,a_r)\,,\\
	S_r(z+a|a,2a,a_3...,a_r) =  \frac{\sqrt{S_r(z|a,a,a_3...,a_r)}}{\sqrt{S_{r-1}(z|2a,a_3...,a_r)}}\,, \\
	S_2(a|a,b)=\sqrt{\frac{b}{a}}\,, \\
	S_2(\frac{a}{2}|a,b)=\sqrt{2}\,, \\
	S_2(\frac{a+b}{2}|a,b)=1\,, \\
	S_1(x|w)=2 \sin(\pi \frac{x}{w})\,.
\end{eqnarray}

In our derivations we also made use of the relation 
\begin{eqnarray} \label{productsine1}
	S_r(Ny|\omega)=\prod_{k_i=0,i=1,...,r}^{N-1} S_r(y+\frac{k_i \omega_i}{N}|\omega)\,,
\end{eqnarray}
and its $y\rightarrow 0$ limit given by 
\begin{eqnarray} \label{productsine2}
	N=\prod_{k_i=0,i=1,...,r,\, \{k \neq 0 \}}^{N-1} S_r(\frac{k_i \omega_i}{N}|\omega)\,.
\end{eqnarray}
Note that both products above run over all $k_i\in\{0,1,\dots,N-1\}$, but in the later the null-vector $k=0$ is excluded.

\section{Outline of the derivation of the partition function for B, C}
\label{secDerivBC}
For the $C_n$ algebras we choose the normalization with the square of the long root being $4$. Then $F_X$ reads 
\begin{eqnarray}
	F_{C_n}= \frac{A_{C_n}}{B_{C_n}} \\
	B_{C_n}=(q^2-1)  \left(q^{2y}-1\right) \\
	A_{C_n}=	(q+1) q^y \left(q^{2 n y}-1\right) \left(q^{2 n y+1}-1\right)+\\
	\left(q^{2 y}-1\right) \left(q^{n y}-1\right) \left(q^{n y+1}-1\right) \left(q^{2 n y+1}-1\right)
\end{eqnarray}

The starting point for the derivation of the partition function is formula (\ref{FintRep}), which we present below in a form more appropriate for transformation into multiple sine functions via expression (\ref{Sint}), i.e.,
\begin{eqnarray}\label{FintRep1}
	\ln Z=-\frac{1}{2}\int_{R_+} \frac{dx}{x} \frac{\left( e^{2xty}-e^{2x\delta}\right)}{\left( e^{2xty}-1\right)\left( e^{2x\delta}-1\right)} F_X(2x,y)\,.
\end{eqnarray}

Next we substitute the expression for $F_X$, and write the numerator of the integrand as a sum of exponents with integer coefficients. Using (\ref{Sint}) we obtain the product/ratio of multiple sine functions in corresponding powers (given by integer coefficients in front of exponents). 
Note that we can add to (\ref{Sint}) (one half of the) residue of the integrand at zero, which is zero. This means that when we transform (\ref{FintRep1}) into terms of multiple sine functions, we can assume that the first terms (Bernoulli polynomials) have been accounted for, since they sum up to zero. We obtain
\begin{eqnarray} \label{ZCinterm1}
	\frac{1}{Z_C}= \\ \nonumber
	\frac{S_4(2y|a)S_4(3y|a)S_4(1+3y|a)S_4(3y+ty|a)S_4(1+3y-y|a)S_4(2+3ty-y|a)}{S_4^2(1+y+2ty|a)S_4(2+y+2ty|a)S_4(4y|a)S_4(y+ty|a)S_4(1+y+ty|a)}  \\ \nonumber
	\times \frac{S_4(2+4ty-y|a)S_4(1+4ty-y|a)S_4(1+ty+3y|a)S_4(2+4ty|a)}{S_2(1+3ty+y|1,2\delta)S_4(2+4ty-2y|a)S_4(y+2ty|a)S_4(2+3ty+y|a)}\,, 
\end{eqnarray}
with parameters of the sine functions given by $a=(2,2y,2ty,2\delta)$, and we introduced Vogel's parameter $t=n+1$ for $C_n$ algebra. 

Next we apply the {\it fusion} property (\ref{fusion}) to pairs (2,3), (5,6), (4,9) and (7,8) in the numerator, and to pairs (1,2), (1,8), (4,5) and (6,9) of sine functions in the denominator. Note that we numerated the sine functions in the numerator, and separately in the denominator, from left to right. In particular, sine function number 1 in the denominator is fused twice, because the corresponding sine function is squared. In all cases, the parameter of the fusion in (\ref{fusion}) is $a=2$. The result of these fusions (in the same order) is given by
\begin{eqnarray}
	\frac{S_4(3y|b)S_4(1+3ty-y|b)S_4(3y+ty|b)S_4(1+4ty-y|b)}{S_4(1+y+2ty|a
		b)S_4(y+ty|b)S_4(1+3ty+y|b)S_4(y+2ty|b)}   \\
	b=(1,2y,2ty,2\delta)
\end{eqnarray}
Further, consider the following two ratios in (\ref{ZCinterm1}): The first sine in the numerator together with  the third one in the denominator, and the 10$th$ sine in the numerator paired with the seventh sine in the denominator. These factors simplify to 
\begin{eqnarray}
	\frac{S_4(2y|a)}{S_4(4y|a)}= S_3(2y|2,2ty,2\delta) \\
	\frac{S_4(2+4ty|a)}{S_4(2+4ty-2y|a)}= \frac{1}{S_3(2+4ty-2y|2,2ty,2\delta) }
\end{eqnarray}
The ratio of the resulting triple sines can be transformed further:
\begin{eqnarray}
	\frac{S_3(2y|2,2ty,2\delta)}{S_3(2+4ty-2y|2,2ty,2\delta) }= \frac{S_2(2+2ty-2y|2,2\delta)}{S_2(2+2ty-2y|2,2ty)}
\end{eqnarray}

Altogether, we obtain for the (inverse of) the partition function:

\begin{eqnarray} \label{ZCinterm2}
	\frac{1}{Z_C} =	  \\ \nonumber
	\frac{S_4(3y|b)S_4(1+3ty-y|b)S_4(3y+ty|b)S_4(1+4ty-y|b)}{S_4(1+y+2ty|b)S_4(y+ty|b)S_4(1+3ty+y|b)S_4(y+2ty|b)} \times \\ \nonumber
	\frac{S_2(2+2ty-2y|2,2\delta)}{S_2(2+2ty-2y|2,2ty)}  \\ \nonumber
	b=(1,2y,2ty,2\delta)
\end{eqnarray}
In this expression we can again numerate the sines separately in the numerator and denominator, and use the fusion formula (\ref{fusion}), with parameter $a=ty$. The pairs to fuse are (1,3) and (2,4) in the denominator, and the same pairs in the numerator. The fusion yields
\begin{eqnarray}
	\frac{S_4(3y|c) S_4(1+3ty-y|c)}{S_4(1+y+2ty|c) S_4(y+ty|c)} \frac{S_2(2+2ty-2y|2,2\delta)}{S_2(2+2ty-2y|2,2ty)}  \\
	c=(1,2y,ty,2\delta)
\end{eqnarray}

Next, we cancel all fourth sines:

\begin{eqnarray}
	\frac{ S_4(1+3ty-y|c)}{S_4(1+y+2ty|c) } = \frac{S_3(1+2ty-y|1,ty,2\delta)}{S_3(1+2ty-y|1,2y,2\delta)} \\
	\frac{S_4(3y|c) }{ S_4(y+ty|c)} = \frac{S_3(y|1,2y,2\delta)}{S_3(y|1,ty,2\delta)}
\end{eqnarray}

Finally, one can cancel all sines with $ty$ in the parameters, and obtain the final answer for the partition function, i.e.,
\begin{eqnarray}
	\frac{1}{Z_C}=	\frac{S_3(y|1,2y,2\delta)}{S_3(1+2ty-y|1,2y,2\delta)} \frac{S_2(2+2ty-2y|2,2\delta)}{S_2(1+ty-y|1,2\delta)}\,.
\end{eqnarray}
Compared with the initial expression, this is a remarkably simple answer. It can be checked to coincide in the unrefined limit ($y=1$) with the partition function given in \cite{SGV}. Another check is the numerical coincidence with the initial expression (\ref{FintRep1}) at random values of the three parameters $\delta, n, y$. 

A similar, although more cumbersome, calculation can be performed for the B case, yielding (\ref{ZBr1}). In particular, we made use of relations (\ref{productsine1}) and (\ref{productsine2}) for the derivation. As before, we confirmed the validity of the final result by a numerical check.

\end{document}